\documentclass[12pt]{article}
\usepackage{amsmath}
\usepackage{graphicx}
\begin{document}

\title{Climate Modeling and Bifurcation}
                   
\author{Mayer Humi\thanks {e-mail: mhumi@wpi.edu.}\\
Department of Mathematical Sciences\\
Worcester Polytechnic Institute\\
100 Institute Road\\
Worcester, MA  01609}

\maketitle
\thispagestyle{empty}

\begin{abstract}

Many papers and monographs were written about the modeling the Earth climate
and its variability. However there is still an obvious need for a module
that presents the fundamentals of climate modeling to students at the 
undergraduate level. The present educational paper attempts to fill in 
this gap. To this end we collect in this paper the relevant climate data and
present a simple zero and one dimensional models for the mean temperature of 
the Earth. These models can exhibit bifurcations from the present Earth 
climate to an ice age or a "Venus type of climate". The models are 
accompanied by Matlab programs which enable the user to  
change the models parameters and explore the impact that these changes 
might have on their predictions on Earth climate.
\end{abstract}

\newpage

\section{Introduction}

The Earth climate and its variations has always been of great interest 
to Humans as it has major impact on Human activities. 
In this paper we present prototype models for the mean temperature of the 
Earth which will enable us to explore possible bifurcations in the Earth 
climate which might due in part to anthropogenic emissions and natural
forcing processes. These model are based in part on the approach presented
in \cite{WS1,GC}. More elaborate and sophisticated models are available in the 
literature \cite{MHS,WGB,GC}.

Much was written lately about the gradual change in the mean temp of  
the earth due to Human intervention which is expected to range by 1-2 degrees
by the end of this century. However the real danger of these changes is that
they may lead to a rapid major change in the Earth climate (viz. climate 
bifurcation \cite{HAD}) in the same way that a rubber band snaps suddenly 
when it is over stretched or a sudden snow avalanche occurs on the slopes of a 
mountain. 

The Earth climate system is composed of the following components: land,
biosphere, atmosphere, ocean and the cryosphere (ice and frozen water),
These components display a broad range of variability on temporal and
spatial scales such as the Dansgaard-Oeschger cycles which occur
quasi-periodically on a millennial time scale or the El Niño-Southern
Oscillation in the equatorial Pacific. Thus a refined model of such
a complex system requires vast amounts of data and elaborate sophistication.
In the following we stick to the basics.

The plan of the is as follows: In Sec 2 we introduce the basic concepts 
needed for climate modeling. Sec $3$ presents the relevant climate data. 
The zero dimensional model and it predictions are discussed in Sec $4$.
The one dimensional climate model is presented in Sec $5$. Sec $6$
discusses some attempts to mitigate the effects of the greenhouse effect.
We end up in Sec $7$ with some conclusions.

\section{Some Basics}

In this section we introduce some basic concepts and data needed
for the modeling of the Earth climate.

\subsection{The Albedo}

When radiation impinges on a (perfect) mirror all of this radiation
is reflected back and the temperature of the mirror remains unchanged.
On the other hand is such radiation impinges on a (perfect) black body
all of the radiation is absorbed by that body. In general however
some of this radiation is absorbed by the body and some is reflected 
back. We say that in this case we are dealing with a "grey body". 
In general the albedo  of a body is the percentage of radiation that 
is reflected back by the body into space. Thus for a perfect mirror the 
albedo is $1$ and for a black body the albedo is zero. For a grey body 
the albedo is a number between zero and one.

The Earth is a grey body. However it albedo changes with time due to 
snow and ice cover, vegetation and Human activities (e.g. 
the paving of asphalt roads)

\subsection{Greenhouse Gases and Clouds}

Radiation from the Sun reaches Earth (mostly) in the visible part of 
the spectrum (viz. wave lengths in the range of 
$380\times 10^{-9}$ - $750\times 10^{-9}$ meters).
Radiation with shorter wavelengths is absorbed (mostly) by the Van-Allen belts.
The Earth absorbs part of this radiation and emit it back in the infra
red part of the spectrum (with wave lengths n the range of 
$7.5\cdot 10^{-7}-1\cdot 10^{-4}$ meters. If the Earth had no atmosphere
this radiation will be reflected into space. However in the Earth atmosphere
some trace gases such as Carbon dioxide ($CO_2$), Methane ($CH_4$) and 
water vapor can absorb this radiation and reflect it back to Earth [4].
This leads to a warming of the Earth surface. (Fig. $1$ presents the 
absorption spectra of several trace gases- source of this figure is unknown).

NOAA Earth System Research Laboratory (Global Monitoring Division)
has been monitoring the concentration of the trace gases in the 
atmosphere from an observation station on Mona Loa mountain in Hawaii
and globally\cite{NOA1,NOA2,NOA3} by averaging these concentrations 
as measured by observation stations throughout the globe. This data is shown in 
Figs. $2,3,4$ (courtesy of "NOAA Global Monitoring Division') they 
demonstrate the almost linear increase in the concentrations of $CO_2$ and 
$CH_4$ over many years which can be attributed Human activities.

The effect of clouds in this scenario is dual. First they block the Sun 
radiation and reflect it into space. At the same time they also reflect
the Earth infra red radiation back to Earth. There is still on going debate 
in the scientific community as to which of these two process is more 
prominent for the determination of the Earth energy balance.

\subsection{\bf Stefan-Boltzmann Law}

Stefan–Boltzmann states that the power radiated from a black body  as a
function of its temperature is given by
$$
P=\sigma A T^4
$$
where $A$ is the body surface area, T is it temperature and
$\sigma=5.67\times 10^{-8}$ $Watts/(m^2 K^4)$ is the Stefan–Boltzmann constant.

For a grey body $P$ has to multiplied by the ``grayness" coefficient of
the body.

\section{A Model for the Mean Temperature of the Earth}

We enumerate here the  data that is needed to model the mean temperature 
of the Earth and then assemble these into model equations.

\begin{enumerate}

\item {\bf The Sun Forcing} 

The rate at which energy from the Sun reaches the top of Earth’s atmosphere 
is called “total solar irradiance” (or TSI)\cite{TSI,CLP,KL}.  
TSI fluctuates slightly 
from day to day and week to week. In addition to these rapid, short-term 
fluctuations, there is an 11-year cycle in TSI measurements related to 
“sunspots” (a part of the Sun’s surface that is temporarily cooler and darker than its neighboring regions).

The evidence shows that although fluctuations in the amount of solar energy 
reaching our atmosphere do influence our climate, the global warming trend 
of the past six decades cannot be attributed to changes in the Sun output.
Fig. 5 which was compiled by Professor G. Kopp\cite{KL} 
(and reproduced here with his permission) is a graph of TSI for the 
last 400 years. It shows clearly 
the "small ice age" during the 17th century. This TSI average is referred to
as the 'solar constant', in power (watts) per square meter. 

From a practical point of view we can consider the Sun rays
arriving in parallel to Earth. This energy flux was measured to 
be $F_s\approx 1362$ $Watt/m^2$ \cite{TSI,CLP,KL}. However the cross section 
of the Earth to this flux is $\pi R^2$ while it surface area is
$4\pi R^2$. Therefore we divide this flux be $4$ when we compute 
the average temperature of the Earth.

The average albedo of the Earth from the upper atmosphere, its planetary 
albedo, is $30$–$35\%$ because of cloud cover, but widely varies locally across 
the surface because of different geological and environmental features.

\item {\bf the Albedo}

The albedo of the earth is constantly changing. We shall model it as having 
two values:
$$
\alpha_M=0.85,\,\,\,\, \alpha_m=0.25;
$$
where $\alpha_M$ is the albedo for Earth covered by snow,ice (frozen Earth),
and $\alpha_m=0.25$ is the albedo for the part of the Earth that is not 
covered by snow, ice etc (i.e not frozen).
The Earth will be considered as frozen if it mean temperature is below
$T_1=240$ Kelvin and unfrozen if it mean temperature is over $T_2=275$ Kelvin.
In between $T_1$ and $T_2$ we use linear interpolation to compute the albedo
\begin{equation}
\label{4.2}
\alpha(T)=\frac{\alpha_m-\alpha_M}{T_2-T_1}(T-T_1)+\alpha_M
\end{equation}
(Observe that the effect of changes in the vegetation of the Earth is not 
taken (directly) into consideration in this model)

More elaborate models for the earth albedo are available\cite{WS2,GC} and 
satellites are being used currently to give accurate real time data on its 
value. It is estimated that the average albedo of the Earth (viz. planetary
albedo) is $30\%$ to $35\%$ because of cloud cover and the effect of 
trace gases, but it varies widely across the surface because of 
different geological and environmental features.

\item{\bf Clouds and the Greenhouse factor}

Let $T_0$ be the ambient temperature for green house effect,
and $\kappa(T)$ the cloud cover due to the green house effect.
(Without the greenhouse effect the mean temperature of the Earth will be
$\approx 240$ Kelvin and effectively there will no cloud cover)

The incoming radiation from the Sun that is absorbed by Earth
is 
$$
R_{in}=\frac{F_sA}{4}(1-\alpha(T))
$$
where $A$ is the surface area of the Earth.
The outgoing infra red radiation from Earth is
$$
R_{out}=\sigma Ag(T) T^4
$$
where $g(t)$ is a lump sum parameter that represents the impact of
the greenhouse gases and clouds. This parameter was modeled (empirically)
by Sellers\cite{WS1} as
$$
g(T)=1-\kappa \tanh[(T/T_0)^6],\,\,\,\, \kappa=0.5
$$
where $T_0\approx 275$ Kelvin.
This expression implies that as $T$ increases $g(T)$ decreases and 
the greenhouse effect becomes stronger ($R_{out}$ decreases)
\end{enumerate}

\section{Zero-dimensional Climate Model}
If we denote by $C$ the heat capacity of the Earth then by the law of 
energy conservation it follows that
\begin{equation}
\label{4.0}
C\frac{dT}{dt}=R_{in}-R_{out}
\end{equation}
at equilibrium (viz. Steady state) $\frac{dT}{dt}=0$ and we must have 
therefore that $R_{in}-R_{out}=0$.

\subsection{Model Predictions}

If the Earth atmosphere was totally transparent with no greenhouse gases
(or if the Earth had no atmosphere) then $g(T)=1$.
The percentage of Sun energy that is reflected by the Earth into space
(viz. albedo) is estimated to be $\alpha(T_E)=0.3$.
The equilibrium mean temperature of the Earth $T_E$ will have to 
satisfy then 
\begin{equation}
\label{4.1}
\frac{F_sA}{4}(1-\alpha(T_E))-\sigma A g(T_E) T_E^4=0
\end{equation}
The solution of this equation yields $T_E=255$ Kelvin which is well below
the freezing point of water. Thus the Earth will be covered by snow and ice.
It follows then that the difference between this value of $T_E$ and the 
(estimated) current mean temperature of $T_0=284.9$ Kelvin is due to the 
greenhouse effect.
To find the value of $g(T)$ that leads to this equilibrium temperature 
we substitute $T_0=284.9$ in (\ref{4.1}) and solve for $g$ 
(with $\alpha(T_E)=0.3$). We find that the current value of $g$ is $0.65$.

To investigate further the insights that can obtained from this model
we wrote a Matlab program $mean03.m$ that implements it.
At first we used this program, with the model parameters that are quoted 
above\cite{WS1,GC}. Using this program with an initial (mean) temperature of 
$285$ Kelvin we obtained Fig. $6$ ($20$ years simulation). 
This initial mean temperature is based on data which was collected by NASA 
which shows that the Earth mean surface temperature in $2017$ was $14.9 C$.
Fig. $6$ shows a steady increase in the average temperature of the 
Earth for the first few years which then stabilizes around $305$ Kelvin.
(This program is available for download from the author web page [14]).

Many of the original parameters\cite{WS1,GC} used in this program are estimates
(at best) and are changing with time. Therefore one can use this program to 
probe also for the dependence (or sensitivity) of the results to the values 
of the various parameters of this model. As an example we varied the value of 
$T_2$ in (\ref{4.2}) from $275$ (original value) to $296.11$ with no
substantial changes in the results. However if we increase $T_2$
to over $296.12$ we obtain Fig $7$ where the Earth mean temperature
plunges to an ice age. Thus the model exhibits a climate bifurcation
if $T_2$ is in between $[296.11,296.12]$.

\section{One-dimensional Climate Model}

In this model the Earth temperature is considered to be a function of time and 
latitude $\theta$. Using this one dimensional model it is possible 
to study the oscillation of the Earth climate between ice ages in the past.

To present this model we need to develop first, two mathematical formulas.

\subsection{Area of a Strip on a Circle}

We want to find a formula for the area of a strip on a circle of radius $R$
that is enclosed between the equator and {\bf south} of latitude line $\theta$ 
(see Fig $8$). This area equals the sum of the area of the rectangle 
$2R\cos\theta\sin\theta$ and the area enclosed by equator and the two arcs 
of the circle. Hence the total area of the strip is
\begin{equation}
\label{6.1}
A(\theta)=2R^2\cos\theta\sin\theta+
2\displaystyle\int_{Rcos\theta}^R \sqrt{R^2-x^2} dx=
R^2(\cos\theta\sin\theta+\theta)=R^2\left(\frac{\sin(2\theta)}{2}+\theta\right).
\end{equation}
It follows that the area of a strip on the circle between $\theta$ and
$\theta+d\theta$ is
\begin{equation}
\label{6.2}
dA=\frac{dA(\theta)}{d\theta}d\theta=R^2(\cos(2\theta)+1)d\theta.
\end{equation}

\subsection{Area of a "polar cap" on a Sphere}

The formula for the area {\bf north} of a line of latitude $\theta$
can be computed directly as in the previous subsection. It is given by,
\begin{equation}
\label{6.3}
S(\theta) = 2\pi R^2(1-\sin(\theta))
\end{equation}
The area enclosed by a strip on the sphere between $\theta$ and 
$\theta+d\theta$ is
\begin{eqnarray}
\label{6.4}
&&\Delta S= 2\pi R^2(1-\sin(\theta))- 2\pi R^2(1-\sin(\theta+\Delta\theta)
\\ \notag
&&=2\pi R^2(\sin(\theta+\Delta\theta)-\sin(\theta))= 
2\pi R^2\frac{\sin(\theta+\Delta\theta)-\sin(\theta)}
{\Delta\theta}\Delta\theta
\end{eqnarray}
which in the limit $\Delta\theta \rightarrow 0$ yields
\begin{equation}
\label{6.5}
dS=2\pi R^2\cos(\theta)d\theta
\end{equation}
\subsection{Steady State 1-dimensional Model}

To derive a one dimensional steady state model for the Earth
temperature as a function of $\theta$ we consider a strip on the Earth 
between $\theta$ and $\theta+d\theta$. The cross section of this strip 
to the sun radiation is given by (\ref{6.2}) . 
The incoming radiation from the Sun that is absorbed
by this strip on Earth is
\begin{equation}
\label{6.6}
R_{in}={F_s}(1-\alpha(T(\theta))dA=
{F_s}(1-\alpha(T(\theta))R^2(\cos(2\theta)+1)d\theta
\end{equation}
The outgoing infra red radiation from Earth from this strip is
\begin{equation}
\label{6.7}
R_{out}=\sigma g(T(\theta) T(\theta)^4 dS=
\sigma g(T(\theta)) T(\theta)^4(2\pi R^2\cos(\theta)d\theta)
\end{equation}
Hence when the system is out of equilibrium and $T=T(t,\theta)$ we have
by the law of energy conservation
\begin{equation}
\label{6.0}
C\frac{\partial T}{\partial t}=k\nabla T+ R_{in}(\theta)-R_{out}(\theta)
\end{equation}
where $R_{in}$ and $R_{out}$ are given by (\ref{6.6}) and (\ref{6.7})
respectively. The term $k\nabla T$ represents the heat transport (or
diffusion) due to temperature difference (viz. gradient) in the
different latitudes (due to atmospheric and ocean currents). 
The coefficient $k$ is called the diffusion coefficient. Without this 
meridional heat transport the equator will become increasingly warmer 
while the poles increasingly colder. An approximation of this term 
(that is used by some authors) is to replace it by a term proportional 
to the temperature difference ${\bar k}(T(t,\theta)-T_{ave})$ where 
$T_{ave}$ is an estimate for the global temperature average.

We observe also that this model equation does not take into account the
change in the Earth tilt with respect to the Sun at different seasons.  

Neglecting the diffusion term in (\ref{6.0}) we must then have at the
steady state $R_{in}=R_{out}$ and the following expression for $T(\theta)$
\begin{equation}
\label{6.8}
{F_s}(1-\alpha(T(\theta))\cos(\theta)=
\pi\sigma g(T(\theta)) T(\theta)^4
\end{equation}

This equation have to be solved either graphically or iteratively since the 
values of $\alpha(T)$ and $g(T)$ depend on the final equilibrium temperature.

The results of this model are rather unusual. If one starts the simulation
from above the freezing temperature (e.g. 280 Kelvin) one obtains Fig $8$
where at least part of Earth is well above the freezing point. On the other 
hand if one starts the simulation from $240$ Kelvin one finds that the Earth
is a frozen ball (see Fig $9$). This demonstrates that this model exhibits 
the existence of two stable climates, one is "warm" and the other is "frozen".
The first corresponds to the present climate while the second corresponds
to an ice age. Matlab programs mod03.m and mod04.m which implement these 
results can be downloaded from \cite{MH}

If the parameters of the system change (e.g if $F_S$ changes) "slightly"
and the Earth is at the warm stable climate it will stay at warm
temperatures. On the other hand if it is in the cold stable climate
it will remain at an ice age. The transition from one stable climate
to another occurs due to variations in the solar output and the $CO_2$
(and other greenhouse gases) concentration cycle in the atmosphere. 

In this model the average Earth temperature as a 
function of the system variables follow a hysteresis diagram and the 
transition between these two stable climates can happen within a very 
short period of time. This is illustrated in Fig. $10$. In this
(``illustrative") plot the red line denote a "warm climate" while 
the blue line "ice age". In the interval $[-0.3849,0.3849]$ both states 
are stable i.e. if the Earth is in the "warm climate" it will remain on 
the same branch if the system parameters change (and vice versa if the 
Earth is an "ice age"). However if the Earth system is in the "warm climate' 
and reachs the hypothetical point $-0.3849$ the "warm climate' destabilizes and the Earth system "jumps" (or transition rapidly viz. bifurcate) to the other 
stable branch which is an "ice age". Similary if the Earth is in an ice age 
and reaches the point $0.3849$ the ice age becames unstable and the Earth 
system transitions to the warm climate. 

\subsubsection{Steady State 1D Model with Diffusion}

In one dimension (and spherical coordinates) the diffusion term for a 
strip between $\theta$ and $\theta+d\theta$ in (\ref{6.0}) has the form
$\frac{k}\nabla T=\frac{k}{R}\frac{dT}{d\theta}\,d\theta$.
Using (\ref{6.6}) and (\ref{6.7}), the steady state equation
(\ref{6.0}) becomes
\begin{equation}
\label{6.9}
{K}\frac{dT}{d\theta}={F_s}(1-\alpha(T(\theta))(1+\cos(2\theta))-
\sigma g(T(\theta)) T(\theta)^4(2\pi\cos(\theta) 
\end{equation}
where $K=\frac{k}{R^3}$. We simulated this equation with $K=1,1.5,2.$
and initial condition $T=280$ Kelvin at the equator ($\theta=0$). 
The solution(s) are depicted in Fig. $11$. Matlab program diffusion.m 
that was used to obtain these results can be downloaded from \cite{MH}

\section{Geoengineering}

Several countries signed the Paris agreement to curb the concentration of
greenhouse gases in the atmosphere. However this agreement is 
"not totally binding" and there is little political will to abide by this 
agreement by some countries.

Several ideas were considered in the last few years to reverse
the possible effects of climate change. One of these is to
release a compound into the stratosphere that would reflect some of the Sun’s
energy back into space.

To experiment and gauge the impacts of such a program Harvard researchers
\cite{HGE,LMR} sent a balloon into the stratosphere, where it
released about 100 grams of calcium carbonate
This compound was chosen since it could stay in the upper
atmosphere for a long period of time and reflect sunlight back into space.

It is expected that this small scale experiment will provide data about
the risks and rewards of a large scale geoengineering programs.

Carbon sequestration is another scheme that was floated as a way to reduce 
greenhouse gas emissions into the atmosphere. Carbon capture and storage 
(CCS) (or carbon capture and sequestration) is the process of capturing 
waste carbon dioxide, transporting it to a storage site, and depositing 
it where it will not enter the atmosphere, normally into underground geological 
formation \cite{BC}. In some cases the captured carbon dioxide is pumped 
underground 
to force more oil out of oil wells. This set-up allows companies to actually 
make money from capturing carbon rather than it just being a financial burden.
Carbon capture has seen the most success in the United States, where so far 
projects have stored nearly 160 million metric tons of carbon dioxide in 
underground geological formations.

\section{Conclusions}

The models presented in this paper are prototype models for the steady state 
temperature of the Earth. They ignore many features that control the Earth's 
climate. In spite of this they provide a "window" to the most 
important factors that influence the Earth climate viz. Albedo and 
greenhouse effects. They provide a warning signal about the possible 
impact of human activities on global warming and climate.

\newpage
\begin{figure}[ht!]
\includegraphics[scale=1,height=160mm,angle=0,width=180mm]{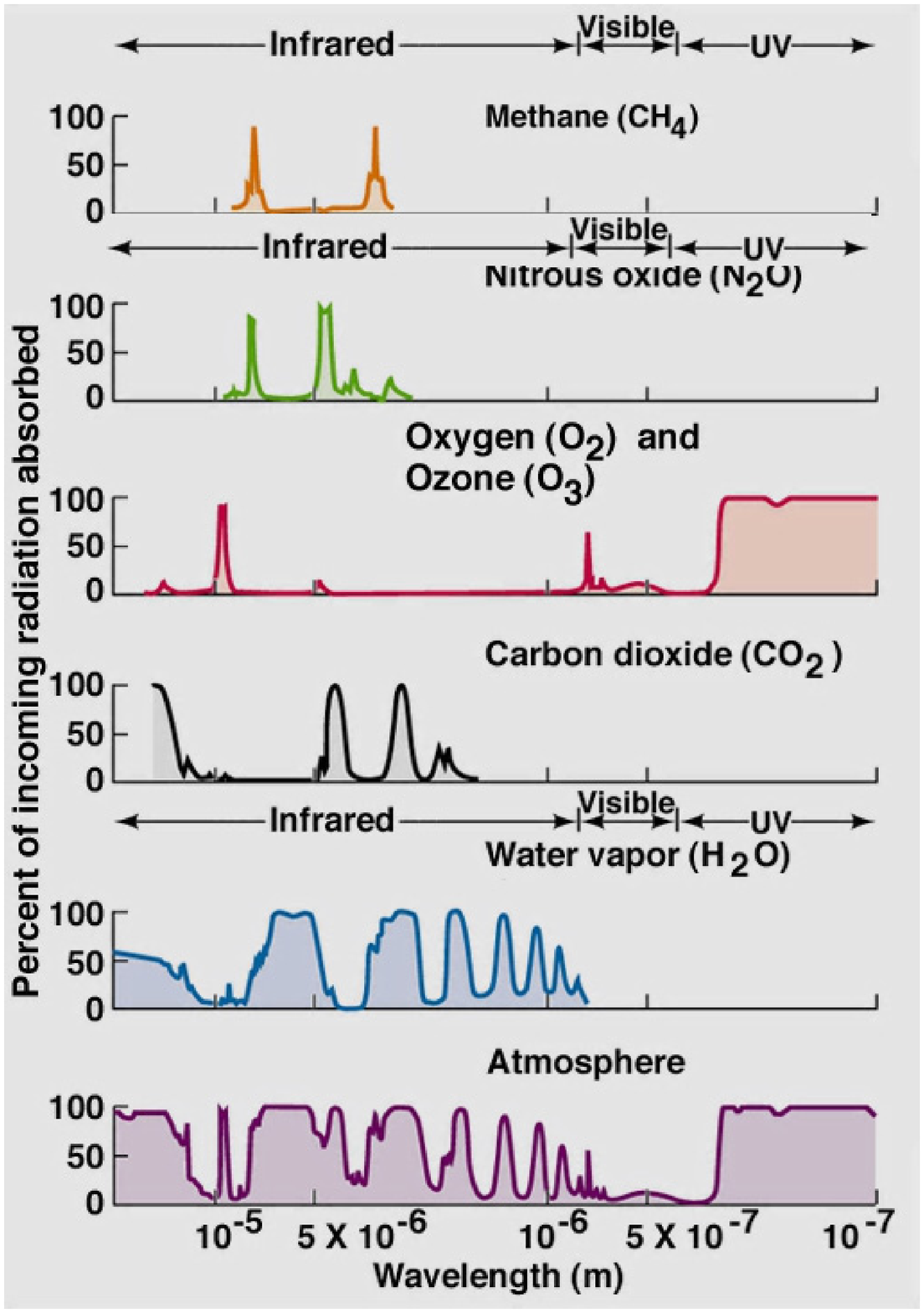}
\label{Figure 1}
\caption{Absorption Spectra of some Trace Gases}
\end{figure}

\newpage
\begin{figure}[ht!]
\includegraphics[scale=1,height=160mm,angle=0,width=180mm]{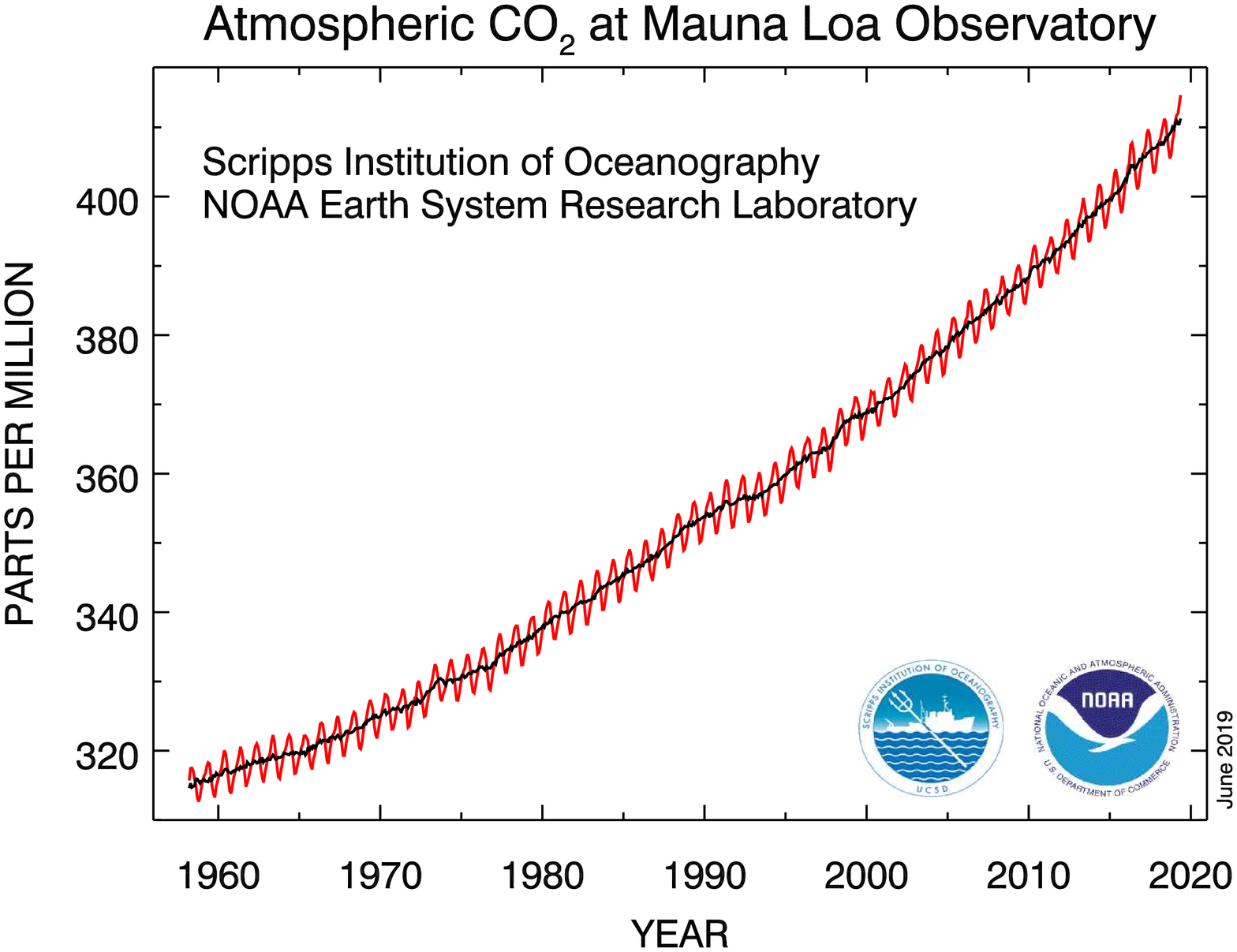}
\label{Figure 2}
\caption{}
\end{figure}
\newpage
\begin{figure}[ht!]
\includegraphics[scale=1,height=160mm,angle=0,width=180mm]{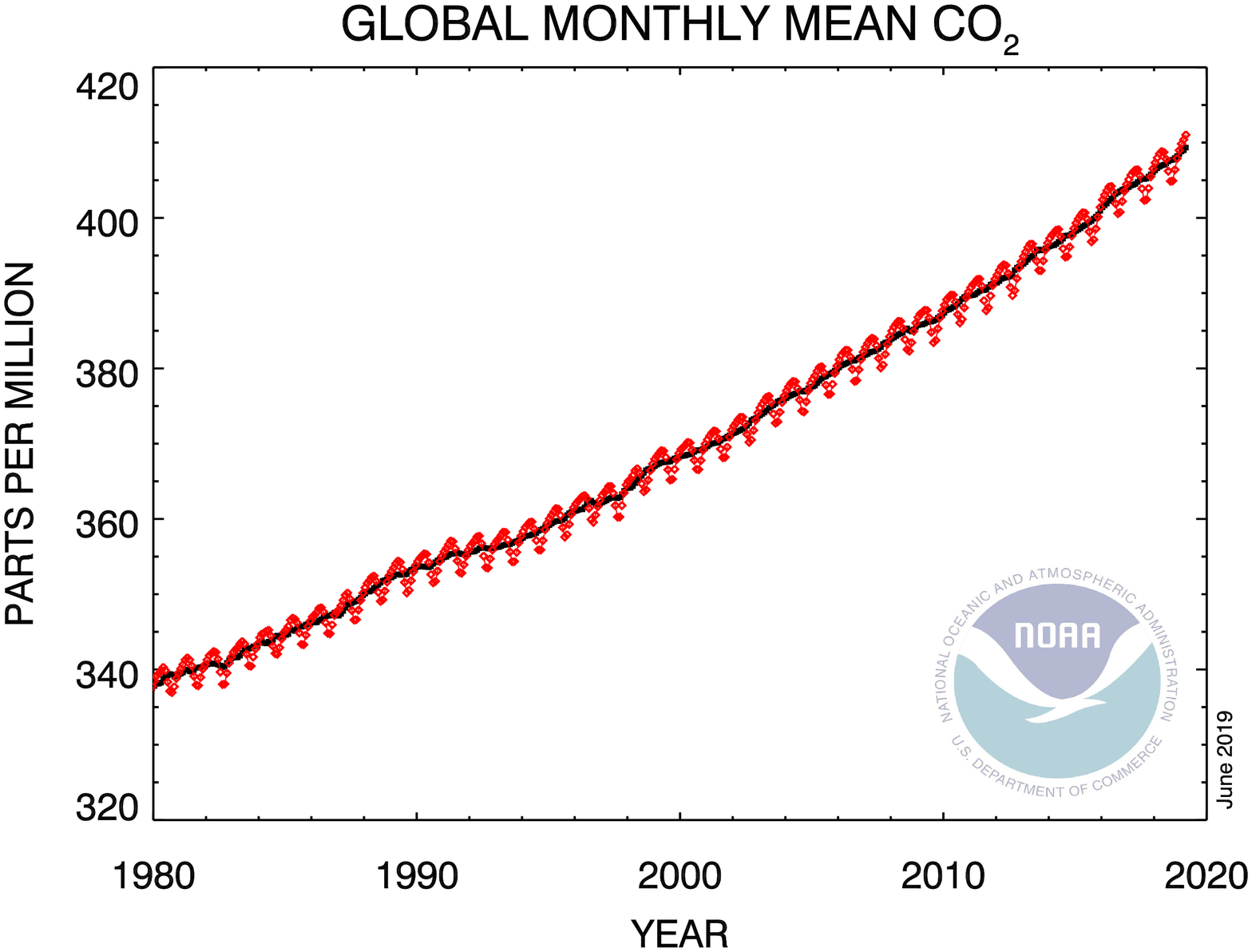}
\label{Figure 3}
\caption{}
\end{figure}
\newpage
\begin{figure}[ht!]
\includegraphics[scale=1,height=160mm,angle=0,width=180mm]{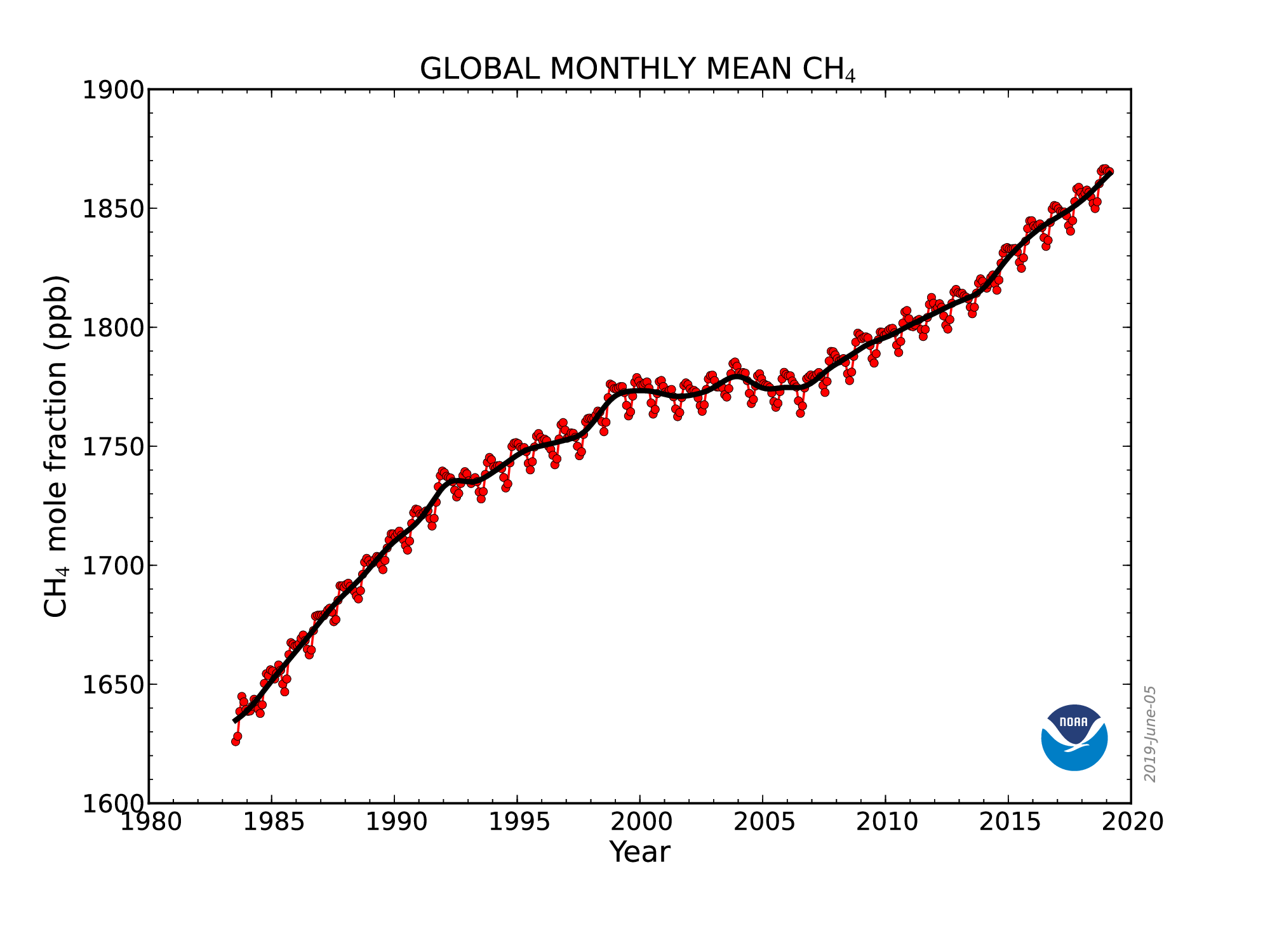}
\label{Figure 4}
\caption{}
\end{figure}
\newpage
\begin{figure}[ht!]
\includegraphics[scale=1,height=160mm,angle=0,width=180mm]{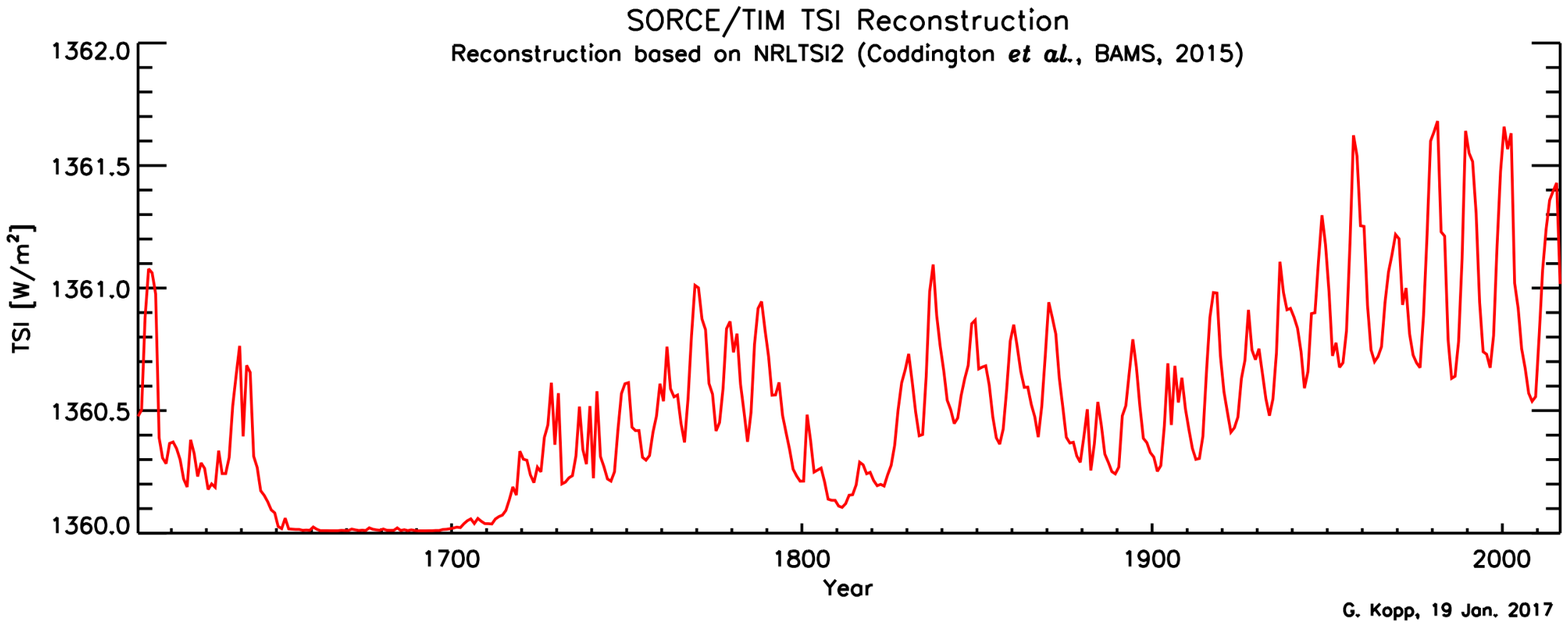}
\label{Figure 5}
\caption{Solar Irradiance}
\end{figure}
\newpage
\begin{figure}[ht!]
\includegraphics[scale=1,height=160mm,angle=0,width=180mm]{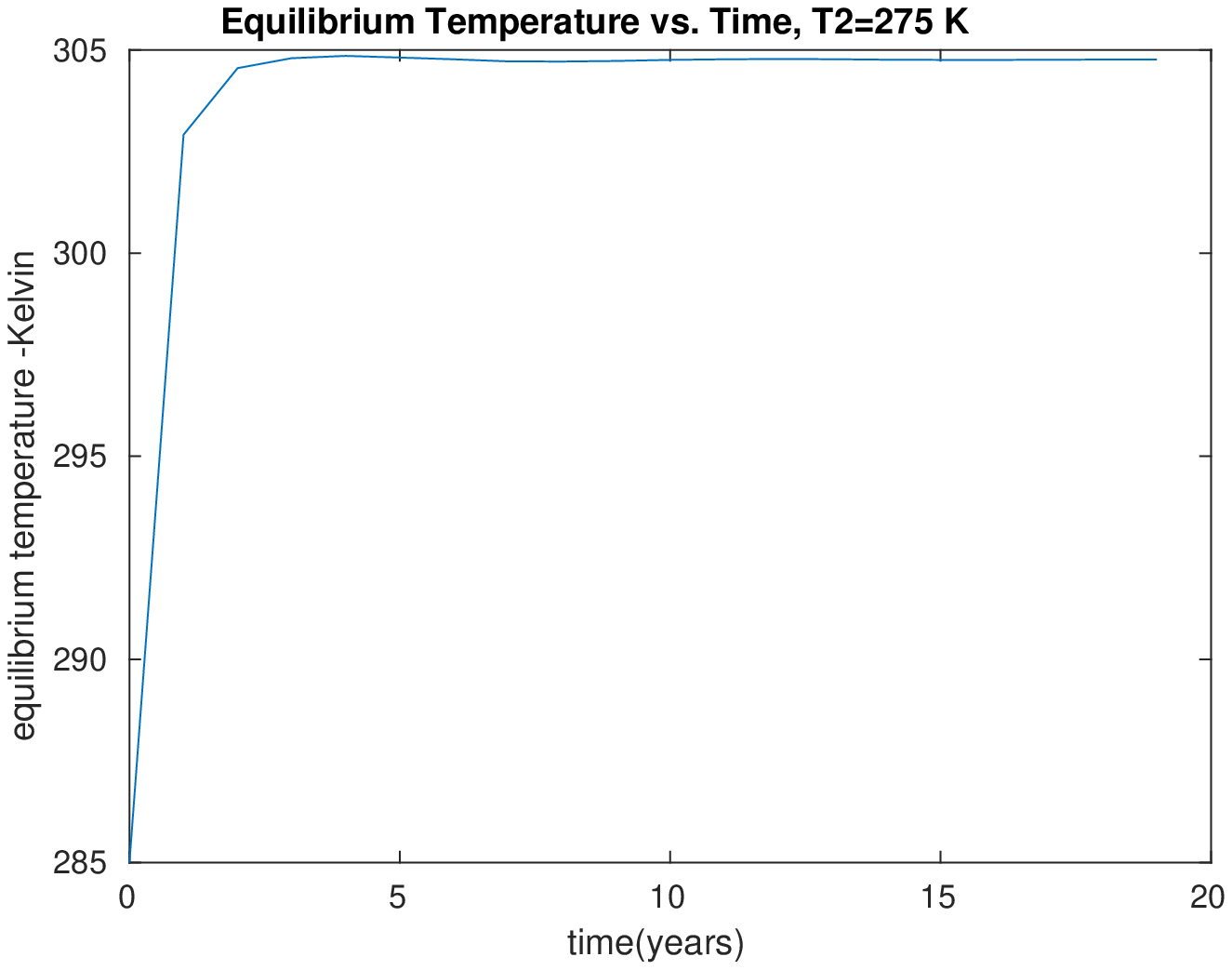}
\label{Figure 6}
\caption{}
\end{figure}
\newpage
\begin{figure}[ht!]
\includegraphics[scale=1,height=160mm,angle=0,width=180mm]{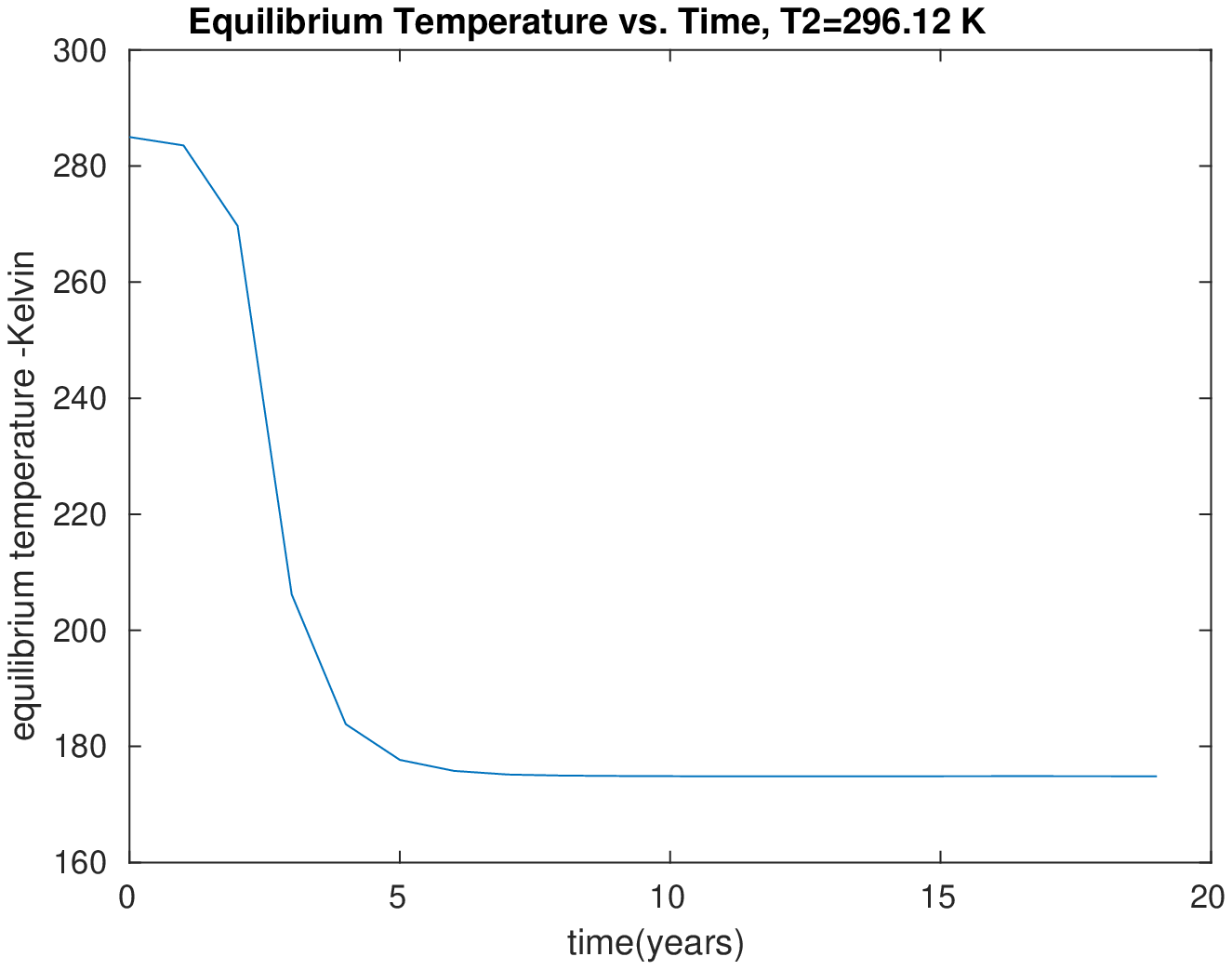}
\label{Figure 7}
\caption{}
\end{figure}
\newpage
\begin{figure}[ht!]
\includegraphics[scale=1,height=160mm,angle=0,width=180mm]{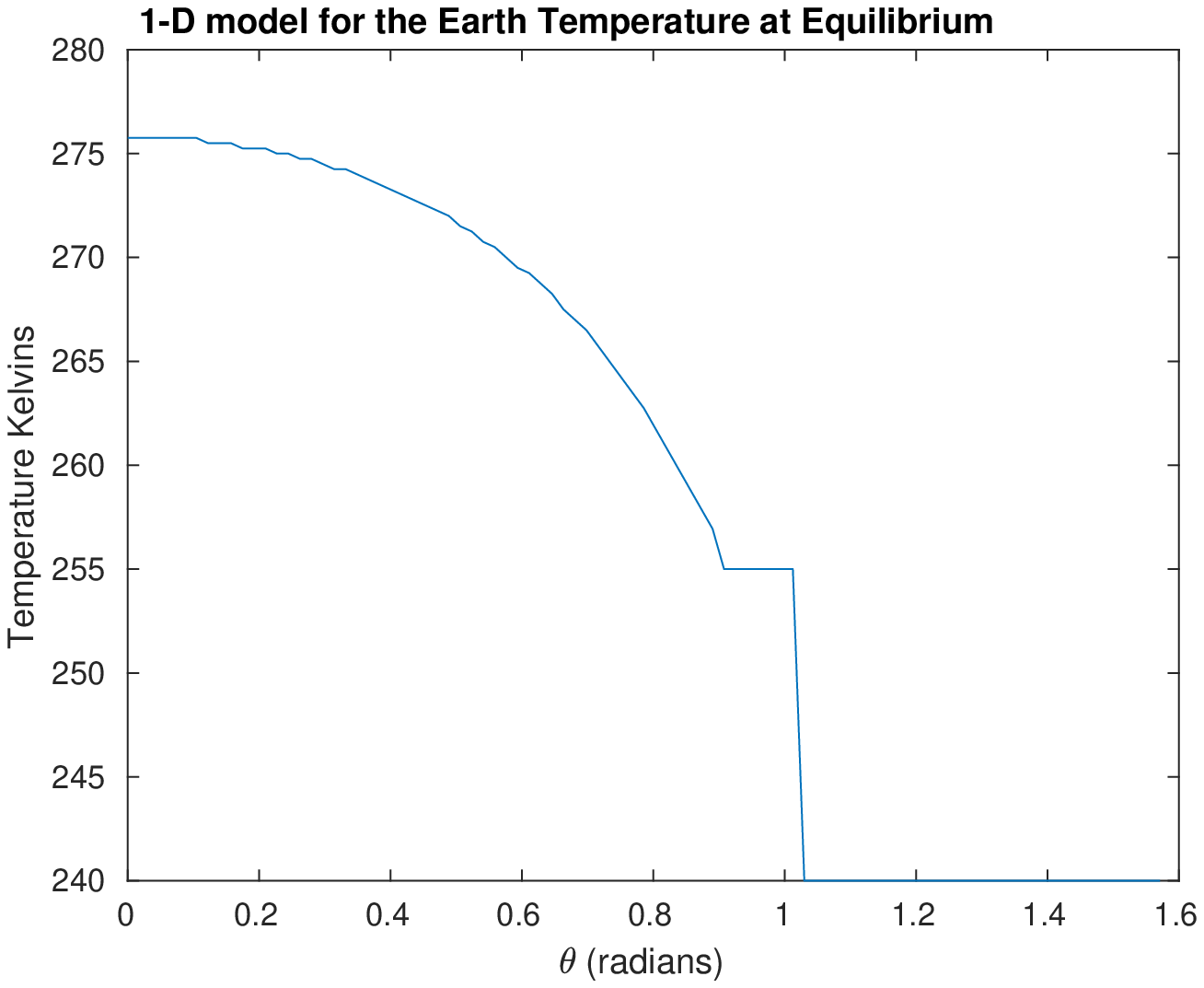}
\label{Figure 8}
\caption{}
\end{figure}
\newpage
\begin{figure}[ht!]
\includegraphics[scale=1,height=160mm,angle=0,width=180mm]{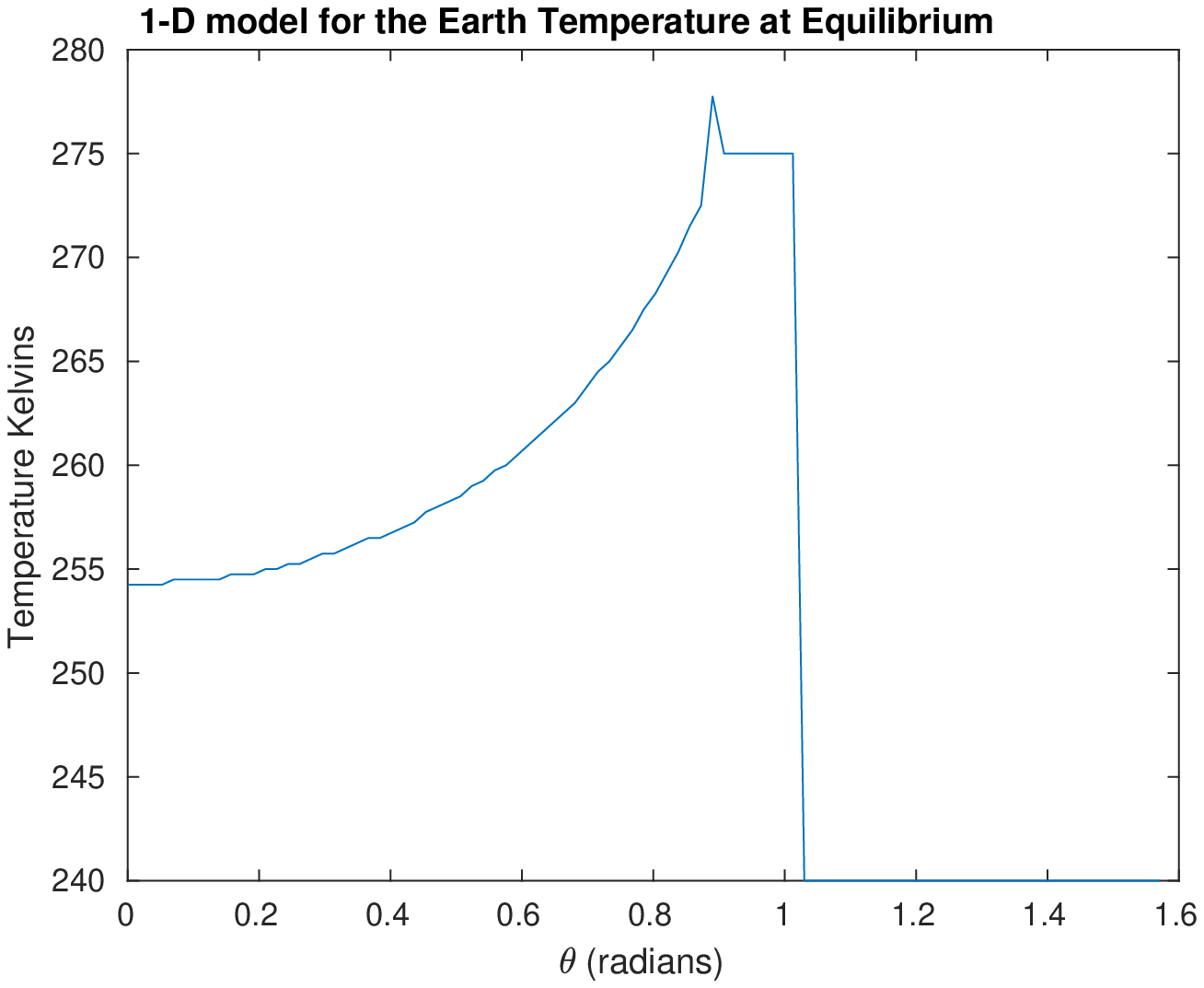}
\label{Figure 9}
\caption{}
\end{figure}
\newpage
\begin{figure}[ht!]
\includegraphics[scale=1,height=160mm,angle=0,width=180mm]{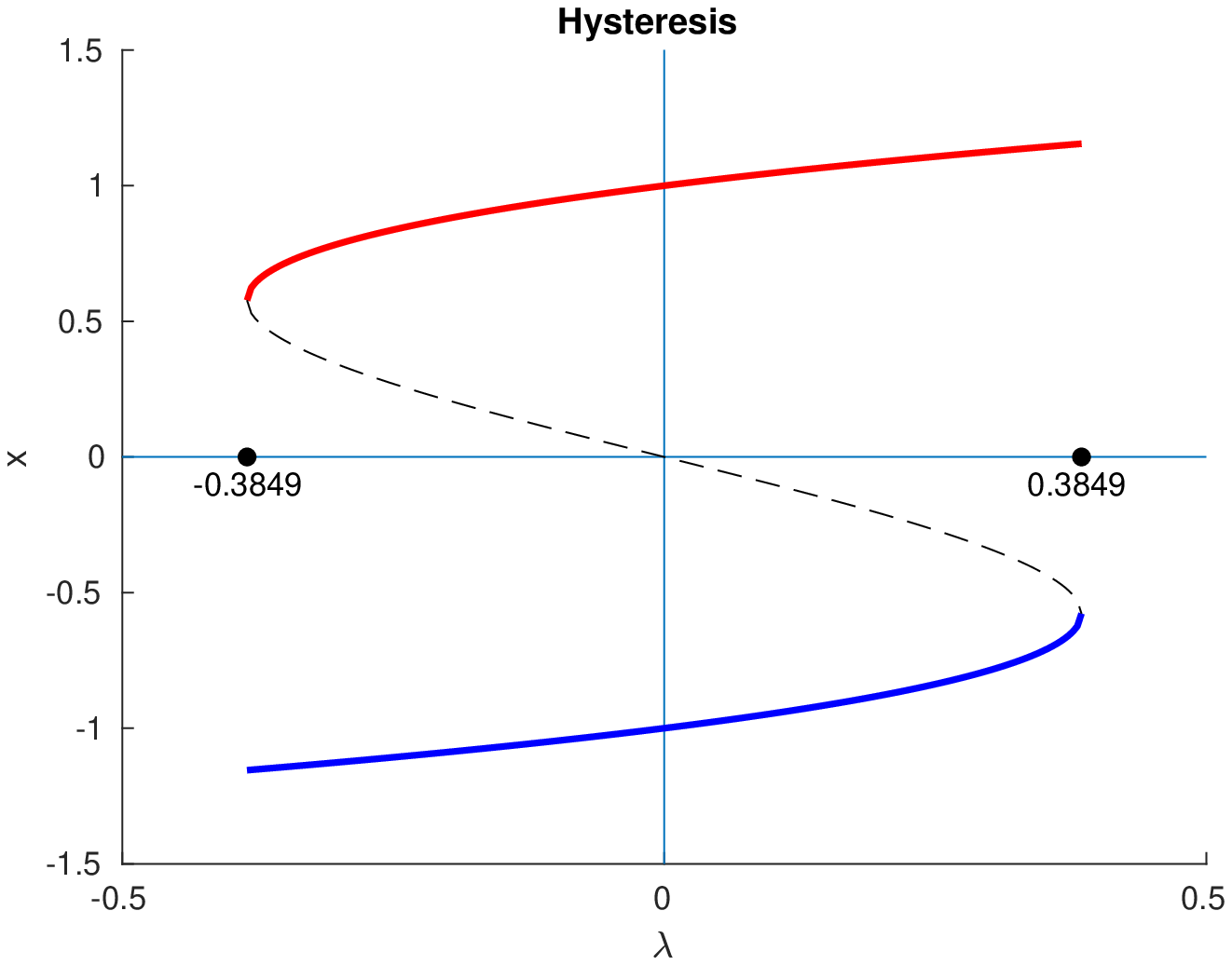}
\label{Figure 10}
\caption{}
\end{figure}
\newpage
\begin{figure}[ht!]
\includegraphics[scale=1,height=160mm,angle=0,width=180mm]{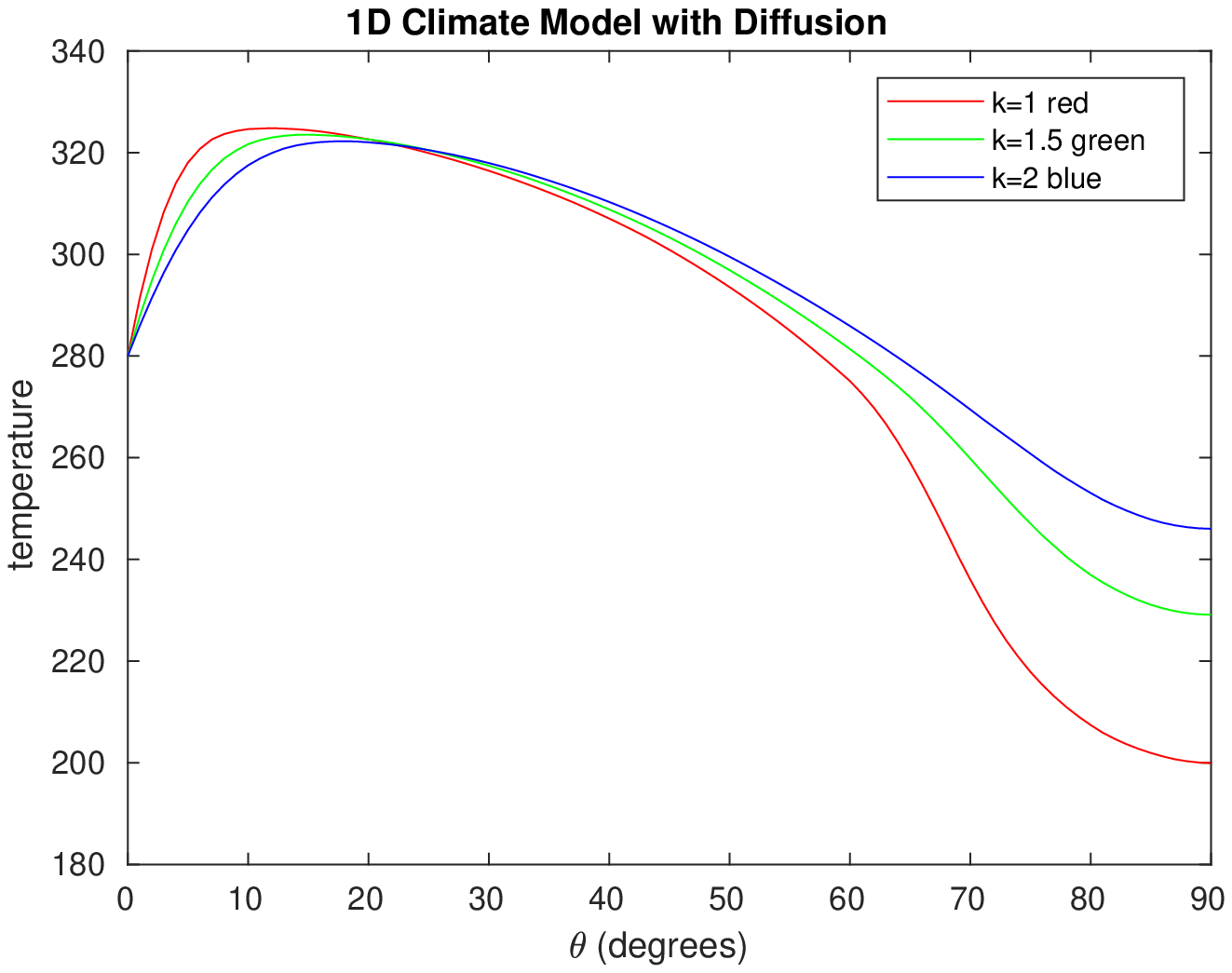}
\label{Figure 11}
\caption{}
\end{figure}
\end{document}